\documentclass[sigconf]{acmart}

\AtBeginDocument{%
  \providecommand\BibTeX{{%
    \normalfont B\kern-0.5em{\scshape i\kern-0.25em b}\kern-0.8em\TeX}}}

\copyrightyear{2023}
\acmYear{2023}
\setcopyright{rightsretained}
\acmConference[KDD '23]{Proceedings of the 29th ACM SIGKDD Conference on Knowledge Discovery and Data Mining}{August 6--10, 2023}{Long Beach, CA, USA}
\acmBooktitle{Proceedings of the 29th ACM SIGKDD Conference on Knowledge Discovery and Data Mining (KDD '23), August 6--10, 2023, Long Beach, CA, USA}
\acmDOI{10.1145/3580305.3599298}
\acmISBN{979-8-4007-0103-0/23/08}

\makeatletter
\gdef\@copyrightpermission{
  \begin{minipage}{0.3\columnwidth}
   \href{https://creativecommons.org/licenses/by/4.0/}{\includegraphics[width=0.90\textwidth]{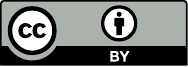}}
  \end{minipage}\hfill
  \begin{minipage}{0.7\columnwidth}
   \href{https://creativecommons.org/licenses/by/4.0/}{This work is licensed under a Creative Commons Attribution International 4.0 License.}
  \end{minipage}
  \vspace{5pt}
}
\makeatother

\usepackage{graphicx}
\usepackage{balance}
\usepackage{microtype}
\usepackage{booktabs}
\usepackage{array}
\usepackage{amsfonts}
\usepackage{amsmath}

\usepackage{amssymb}
\usepackage{multirow}
\usepackage{multicol}
\usepackage{listings}
\usepackage{subfigure}
\usepackage{graphicx}
\usepackage{enumitem}
\usepackage{hyperref}
\usepackage{xcolor,soul}

\newtheorem{problem}{Problem}
\newtheorem{defn}{Definition}
\newtheorem{obs}{Observation}
\begin{document}

\def\blue{\textcolor{blue}}

\title{DECOR: Degree-Corrected Social Graph Refinement for Fake News Detection}

\author{Jiaying Wu}
\affiliation{
  \institution{National University of Singapore}
  \country{}
  }
\email{jiayingwu@u.nus.edu}
\author{Bryan Hooi}
\affiliation{
  \institution{National University of Singapore}
  \country{}
  }
\email{bhooi@comp.nus.edu.sg}

\begin{abstract}

Recent efforts in fake news detection have witnessed a surge of interest in using graph neural networks (GNNs) to exploit rich social context. Existing studies generally leverage fixed graph structures, assuming that the graphs accurately represent the related social engagements. However, \emph{edge noise} remains a critical challenge in real-world graphs, as training on suboptimal structures can severely limit the expressiveness of GNNs. Despite initial efforts in graph structure learning (GSL), prior works often leverage node features to update edge weights, resulting in heavy computational costs that hinder the methods' applicability to large-scale social graphs. In this work, we approach the fake news detection problem with a novel aspect of \emph{social graph refinement}. We find that the \emph{degrees} of news article nodes exhibit distinctive patterns, which are indicative of news veracity. Guided by this, we propose DECOR, a novel application of Degree-Corrected Stochastic Blockmodels to the fake news detection problem. Specifically, we encapsulate our empirical observations into a lightweight social graph refinement component that iteratively updates the edge weights via a learnable degree correction mask, which allows for joint optimization with a GNN-based detector. Extensive experiments on two real-world benchmarks validate the effectiveness and efficiency of DECOR. \footnote{Data and code are available at: \url{https://github.com/jiayingwu19/DECOR}.}

\end{abstract}



\maketitle

\begin{figure}[t]
    \centering
    \includegraphics[width=\columnwidth]{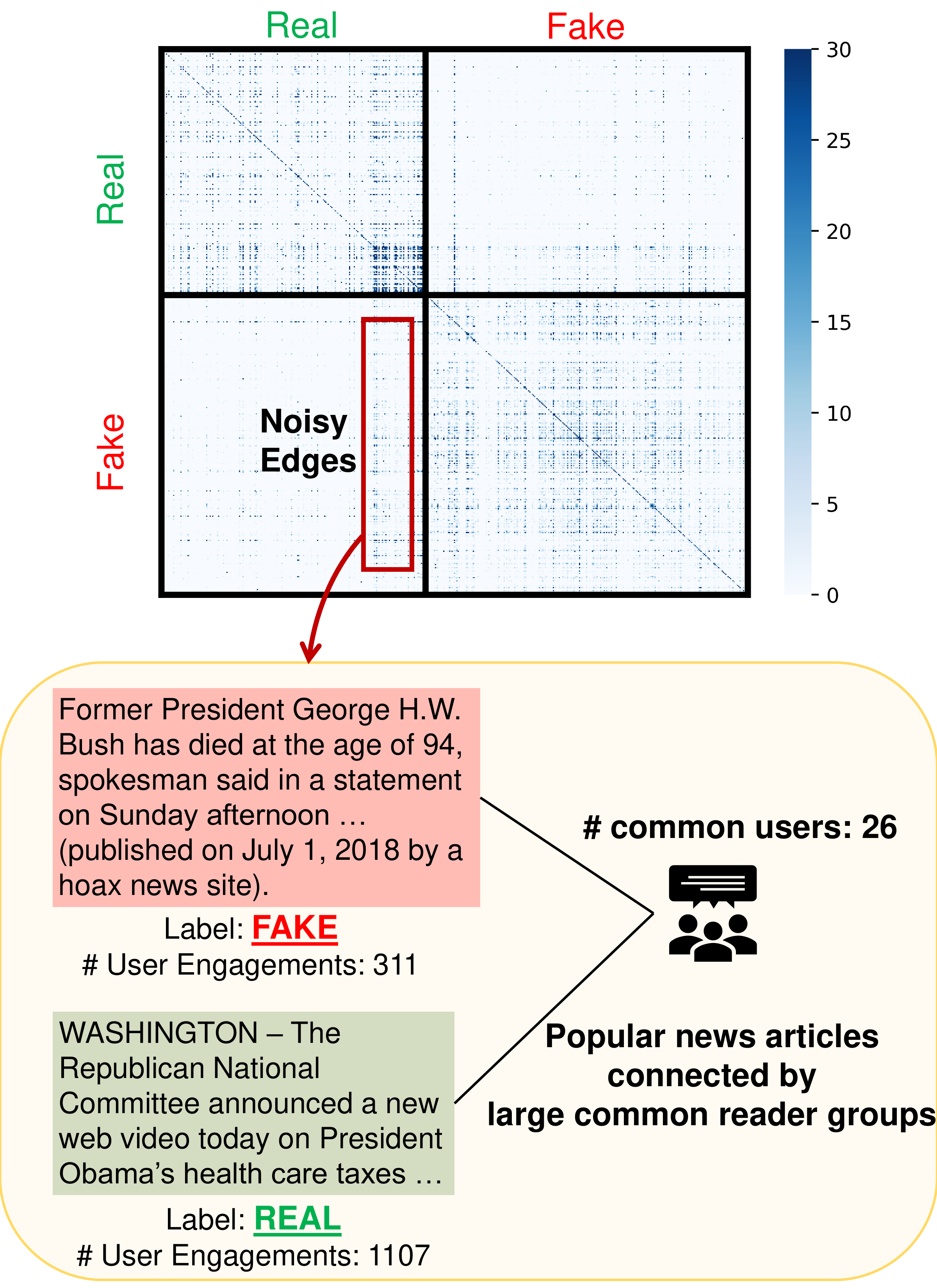}
    \caption{A motivating example for social graph refinement. Darker spots represent a larger number of common readers between two news articles. Weights are clipped at $30$ for a clear visualization.}
    \label{fig:noise-example}
\end{figure}

\section{Introduction}
\label{sec:intro}

Automated detection of fake news stories containing intentionally distorted facts is a major focus of public discourse and scientific research \cite{conroy2015auto,cui2020deter,hu2021compare,ruchansky2017csi,shu2019defend}. Drawing inspiration from the expressive Graph Neural Networks (GNNs) \cite{kipf2017semi,hamilton2017inductive,xu2018how}, a substantial body of research incorporates graphs with rich social context \cite{min2022divide,nguyen2020fang,shu2019beyond} to encode the news dissemination patterns and user responses. Despite varying choices of feature types and GNN backbones, existing approaches are relatively consistent in the design of social graphs. More specifically, the graphs typically contain social users and news articles, which allow GNNs to leverage the relations between structural patterns and news veracity (e.g., closely connected nodes tend to have similar preferences or veracity-related properties). This facilitates the aggregation and propagation of crowd wisdom among connected articles and users, yielding more accurate predictions. Therefore, existing works typically consider social graphs as a high fidelity representation of the social context, the structure of which is kept unchanged throughout model training. 

However, we find that \textit{noisy edges} remain an inevitable challenge for learning on social graphs. A prominent cause is that influential news articles tend to share a large number of common readers (i.e. social users), regardless of their veracity labels. In Figure \ref{fig:noise-example}, we illustrate one such case via visualizing the adjacency matrix of a graph constructed with news articles, termed as the news engagement graph (detailed formulation relegated to Section \ref{sec:newsgraph}). Here, the edge weights are positively correlated with the number of common readers, and larger edge weights imply closer connections between news articles. Figure \ref{fig:noise-example} shows that the largest weights are assigned to the diagonal, and areas representing real news pairs and fake news pairs are also darker. This is expected, given the self-loops and frequent interactions within groups of real news and fake news spreaders. However, the figure also shows scattered dark spots representing noisy edges between real news and fake news. We observe that \textbf{edge noise is degree-related}, in that large edge weights are often distributed along certain rows and columns.

Noisy edges severely undermine the effectiveness of GNNs, as the message passing mechanism \cite{gilmer2017neural} propagates noise and contaminates node representations. However, little effort has been made to mitigate this issue in the fake news detection scenario. Despite preliminary efforts of graph structure learning (GSL) methods in denoising edges for real-world graphs (e.g. citation networks) \cite{dai2022towards,jin2020graph}, existing GSL methods cannot be readily applied to fake news detection, as they generally leverage pairwise node feature similarity to guide edge weight updates. Given the large scale of social graphs, similarity-guided GSL becomes less feasible and raises critical deployment challenges.

In this work, we investigate the fake news detection problem from a novel aspect of \textbf{social graph refinement}. Given a set of news articles, we construct and refine a news engagement graph that connects the articles with common readers. Guided by our observation of degree-related edge noise, we explore veracity-related degree patterns on the news readership graph, and make two key findings: \textit{\textbf{(1)}} nodes representing fake news and real news exhibit distinctive degree distributions; and \textit{\textbf{(2)}} grouping edges by the veracity labels of the articles they connect, different edge groups demonstrate a clear difference regarding the relationship between degrees and the number of common readers. 

Motivated by our empirical findings on veracity-related degree and co-engagement patterns, we present \underline{\textbf{De}}gree-\underline{\textbf{Cor}}rected Social Graph Refinement (DECOR), a novel social graph refinement framework for fake news detection. DECOR is based on a flexible extension of the Degree-Corrected Stochastic Blockmodel (DCSBM) \cite{karrer2011stochastic}, a graph generative model that allows us to simultaneously consider the effects of degree and node labels, in a tractable probabilistic manner. DECOR suppresses noise in the news engagement graph by downweighting the noisy edges, specifically via learning a \textit{social degree correction mask} based on a theoretically motivated likelihood ratio-based statistic under the DCSBM model, with a nonlinear relaxation to improve the flexibility of the model.
DECOR utilizes the degree correction mask to adjust the edge weights of the news engagement graph, which is then jointly optimized with a GNN-based classifier to predict news veracity. 

In summary, our contributions are as follows:
\begin{itemize} [leftmargin=*]
\item \textbf{Empirical Findings}: We present two novel findings, on how both \emph{degree} and \emph{co-engagement} closely relate to news veracity.
\item \textbf{Principled DCSBM-based GSL}: Motivated by our empirical findings, we propose DECOR, a GSL approach for reducing edge noise, based on a theoretically motivated \emph{likelihood ratio}-based statistic under the DCSBM model, combined with a nonlinear relaxation. 
\item \textbf{Efficiency}: Unlike existing GSL approaches, DECOR avoids using high dimensional features as input for GSL, and is also linear in the number of edges, thus being 7.6 - 34.1 times faster than existing GSL approaches.
\item \textbf{Effectiveness}: DECOR improves F1 score by 4.55\% and 2.51\% compared to the best baseline on two real-world fake news detection benchmarks, consistently improves the performance of multiple GNN baselines in a plug-and-play manner, and outperforms baselines under label scarcity.
\end{itemize}

\section{Related Work}

\subsection{Fake News Detection}

Fake news detection is commonly considered as a binary classification problem, with the goal of accurately predicting a given news article as real or fake. Among existing studies, \textbf{content-based methods} extract semantic patterns from the news content using a wide range of deep learning architectures that include RNNs \cite{przybyla2020capturing}  and pre-trained language models (PLMs) \cite{li2020connecting,pelrine2021surprising}. Some methods also guide model prediction with auxiliary information including knowledge bases \cite{cui2020deter,dun2021kan,hu2021compare,wu2021incorp},  evidence from external sources \cite{chen2022evidence,sheng2021inter,xu2022evidence}, visual information \cite{chen2022cross,shang2022duo,wang2021multimodal,zhou2020safe}, and signals from the news environment \cite{sheng2022zoom}. As fake news detection is often deeply rooted in the social context, \textbf{propagation-based methods} incorporate various social features including user responses and opinions \cite{ruchansky2017csi,monti2019fake,shu2019beyond,shu2019defend, yang2022rein,yang19unsupervised}, user-user following relations \cite{min2022divide}, news sources \cite{nguyen2020fang}, and user history posts \cite{dou2021user} to guide model prediction. Despite the rich social information incorporated, little effort has been made to explore direct relations between news articles and the properties of veracity-related news-news connections. Moreover, many methods are vulnerable to structural noise in social graphs, as they typically adopt fixed graph structures during training.

\subsection{Structure Learning for Robust GNNs}

Graph Neural Networks (GNNs) have demonstrated impressive potential in learning node and graph representations \cite{kipf2017semi,hamilton2017inductive,morris2019graphconv,xu2018how}. Despite the prior success, extensive studies have demonstrated that GNNs are highly vulnerable to adversarial attacks in terms of structural noise \cite{dai2018adversarial, wang2019attacking,zugner2018adversarial}. To alleviate this issue, numerous works have focused on learning optimized structures for real-world graphs, specifically via edge denoising \cite{entezari2020all,tang2020transfer,jin2020graph,dai2022towards,wu2019adversarial}. Motivated by the observation that noisy edges connect nodes with dissimilar features \cite{entezari2020all}, existing methods are generally guided by feature similarity measures. For instance, \cite{wu2019adversarial} conducts edge pruning based on the Jaccard similarity between paired node features, Pro-GNN \cite{jin2020graph} employs the feature smoothness regularization alongside low-rank constraints, and RS-GNN \cite{dai2022towards} utilizes node feature similarity to guide the link prediction process. Nevertheless, graph structure learning (GSL) remains underexplored under the social context of fake news detection. Existing GSL methods are not readily applicable to this task, given the high computational costs incurred in computing pairwise similarity measures between high-dimensional news article representations on large-scale social graphs. While initial efforts have been made in conditioning the edge metric with node degrees for coordination detection \cite{zhang2021vigdet}, the fixed adjustment formula adopted by existing work cannot fully capture the complex relations between degree-related properties, which may vary greatly across datasets. To the best of our knowledge, we propose the first learnable framework for social graph refinement, which leverages low-dimensional degree-related properties to flexibly adjust the edge weights of a news engagement graph for enhanced fake news detection.

\section{Preliminary Analysis}
\label{sec:prelim}

In this section, we formally define the fake news detection problem, establish a social context graph that encodes user engagements in disseminating news articles, and conduct preliminary analysis to explore the veracity-related structural patterns. 

\subsection{Problem Formulation}
\label{sec:formulation}

Let $\mathcal{D}$ be a fake news detection dataset containing $N$ samples. In the social media setting, we define the dataset as 
$$\mathcal{D}=\{\mathcal{P},\mathcal{U},\mathcal{R}\},$$
where $\mathcal{P}=\{p_1,p_2,\dots,p_N\}$ is a set of questionable \textbf{news articles}, $\mathcal{U}=\{u_1,u_2,\dots\}$ is a set of related \textbf{social users} who have spread at least one article in $\mathcal{P}$ via reposting on social media. $\mathcal{R}$ represents the set of \textbf{social user engagements}, in which $r \in \mathcal{R}$ is defined as a triple $\{(u,p,k)|u\in \mathcal{U}, p\in \mathcal{P} \}$ (i.e. user $u$ has given $k$ responses to the news article $p$ in terms of \textit{reposts}). In line with most existing studies, we treat fake news detection on social media as a binary classification problem. Specifically, $\mathcal{P}$ is split into training set $\mathcal{P}_{train}$ and test set $\mathcal{P}_{test}$. Article $p\in\mathcal{P}_{train}$ is associated with a ground-truth label $y$ of $1$ if $p$ is fake, and $0$ otherwise. We formulate the problem as follows: 
\begin{problem}
[Fake News Detection on Social Media] Given a news dataset $\mathcal{D}=\{\mathcal{P},\mathcal{U},\mathcal{R}\}$ and ground-truth training labels $\mathcal{Y}_{train}$, the goal is to learn a classifier $f$ that, given test articles $\mathcal{P}_{test}$, is able to predict the corresponding veracity labels $\mathcal{Y}_{test}$. 
\end{problem}

\subsection{News Engagement Graph}
\label{sec:newsgraph}
The positive correlation between social user preferences and the user's news consumption habits has been acknowledged by prior research \cite{bakshy2015exposure}. Specifically, social media creates an \textit{echo chamber}, where individual beliefs can be continuously reinforced by communication and repetition within like-minded social groups \cite{garimella18political}. 

Motivated by this, we propose to capture the news veracity signals embedded in social user engagements. To distill a comprehensive representation of user preferences, we set a threshold to filter the users with less than $3$ engagements with news articles, and focus on a subset $\mathcal{U}_{A} \subset \mathcal{U}$ containing active users.
Specifically, we construct a \textit{user engagement matrix} $\mathbf{E} \in \mathbb{R}^{|\mathcal{U}_A|\times N}$. Element $\mathbf{E}_{ij}$ represents the number of interactions between user $u_i$ and news article $p_j$, the value of which is retrieved from the corresponding entry $(u_i,p_j,k_{ij})\in\mathcal{R}$. 

Given the news consumption patterns of active social users, we further propose to link the news articles that attract similar user groups via constructing an weighted undirected \textit{news engagement graph} $\mathcal{G}=\{\mathcal{P},\mathcal{E}\}$. The adjacency matrix $\mathbf{A}\in\mathbb{R}^{N\times N}$ of $\mathcal{G}$ is formulated based on overlapping user engagement patterns in $\mathbf{E}$, specifically as:
\begin{equation}
    \mathbf{A}=\mathbf{E}^\top \mathbf{E}. 
\label{eq:adj-formulation}
\end{equation}

Intuitively, element $\mathbf{A}_{nk}$ in $\mathbf{A}$ can be interpreted as the number of 2-hop paths (i.e., news - user - news) between two news articles $p_n$ and $p_k$. Hence, a larger $\mathbf{A}_{nk}$ value represents stronger common interest between the reader groups of news article, implying shared opinions or beliefs in the users' news consumption preferences. 

\begin{figure}[t]
    \centering
    \subfigure[\textbf{PolitiFact}]{\includegraphics[width=0.49\columnwidth]{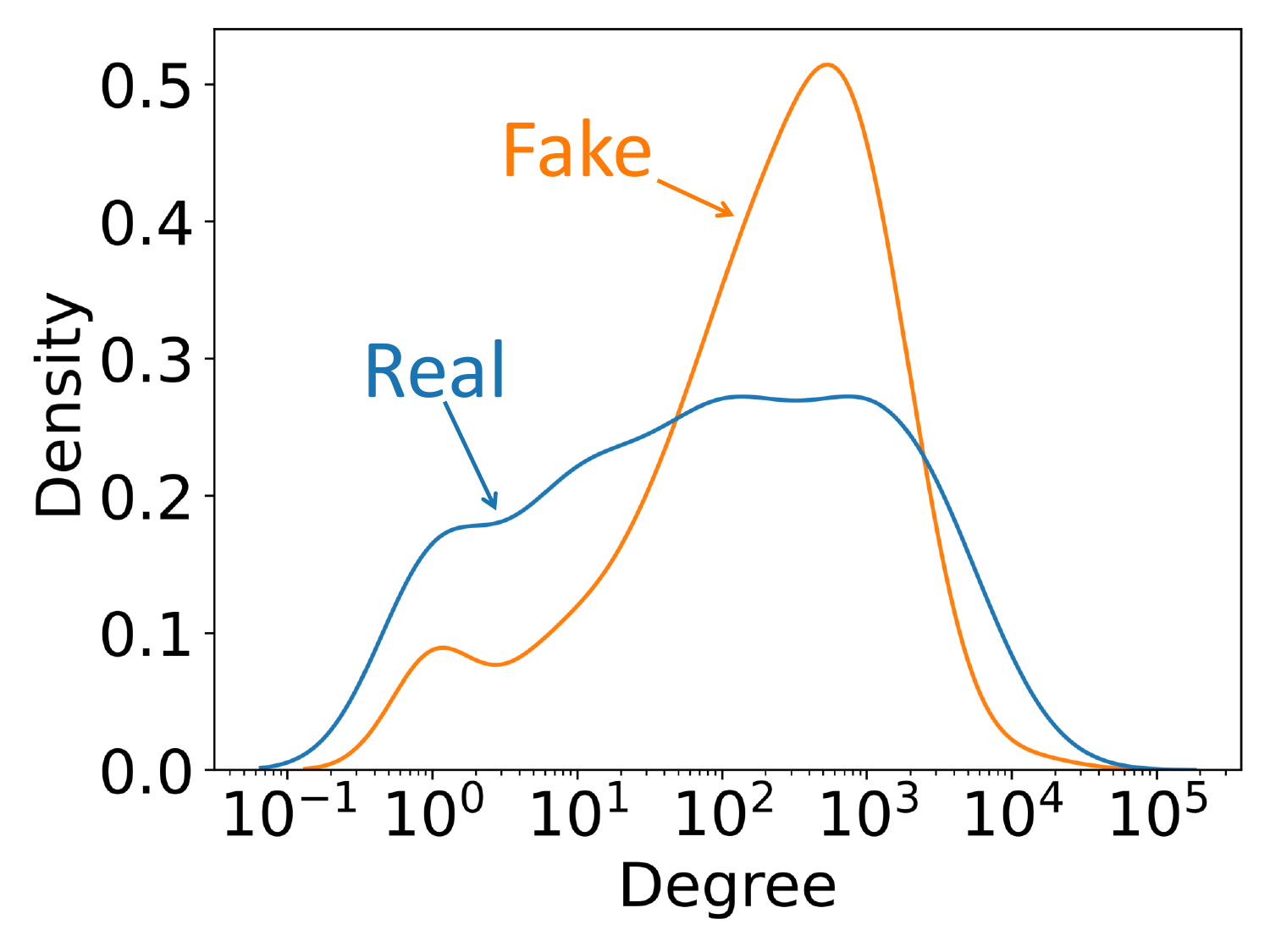}}
    \subfigure[\textbf{GossipCop}]{\includegraphics[width=0.49\columnwidth]{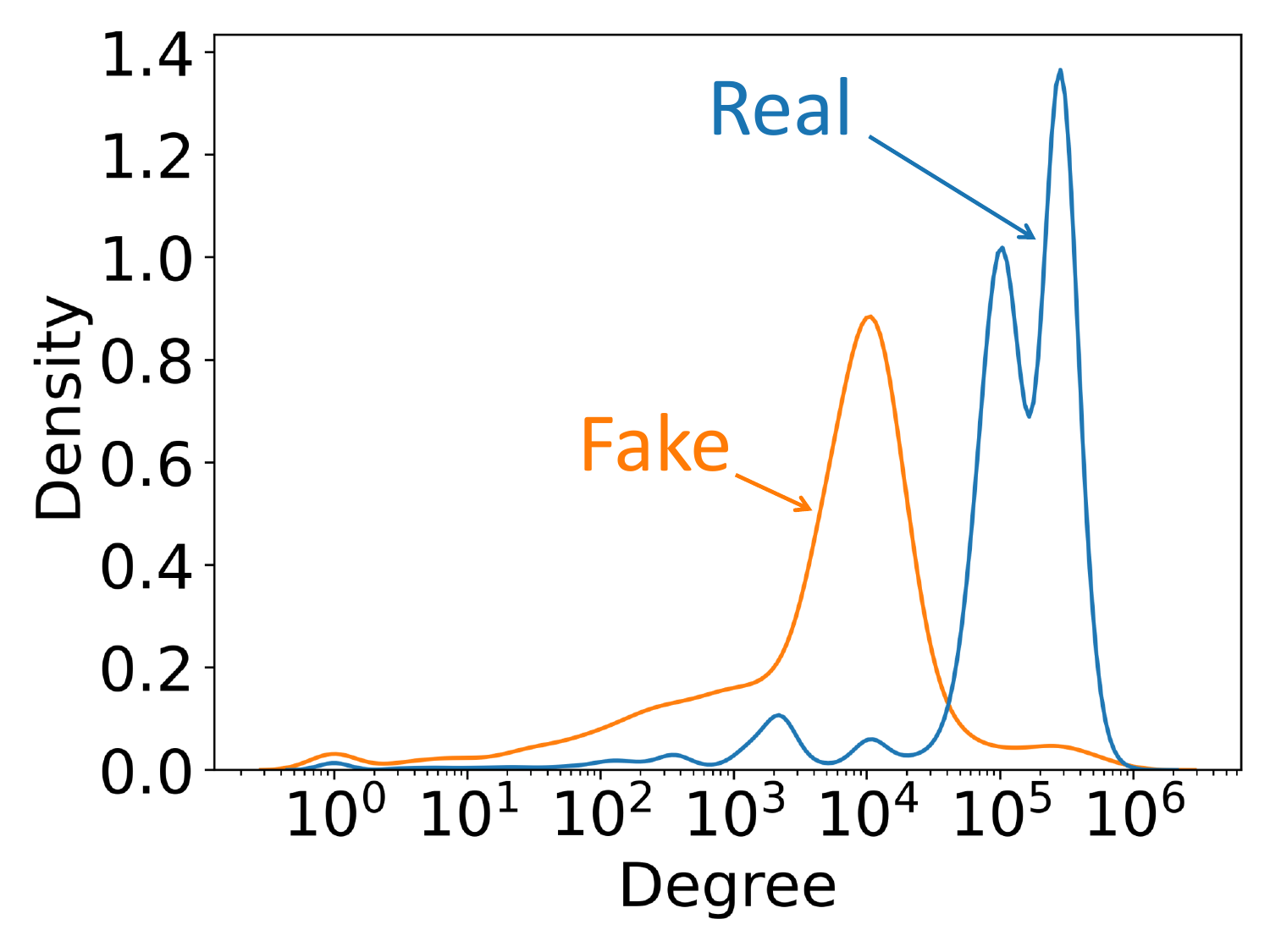}}
    \caption{KDE plot of node degree distributions on the news engagement graph.}
    \label{fig:degree-cred}
\end{figure}

\begin{figure}[t]
    \centering
    \subfigure[\textbf{PolitiFact}]{\includegraphics[width=0.49\columnwidth]{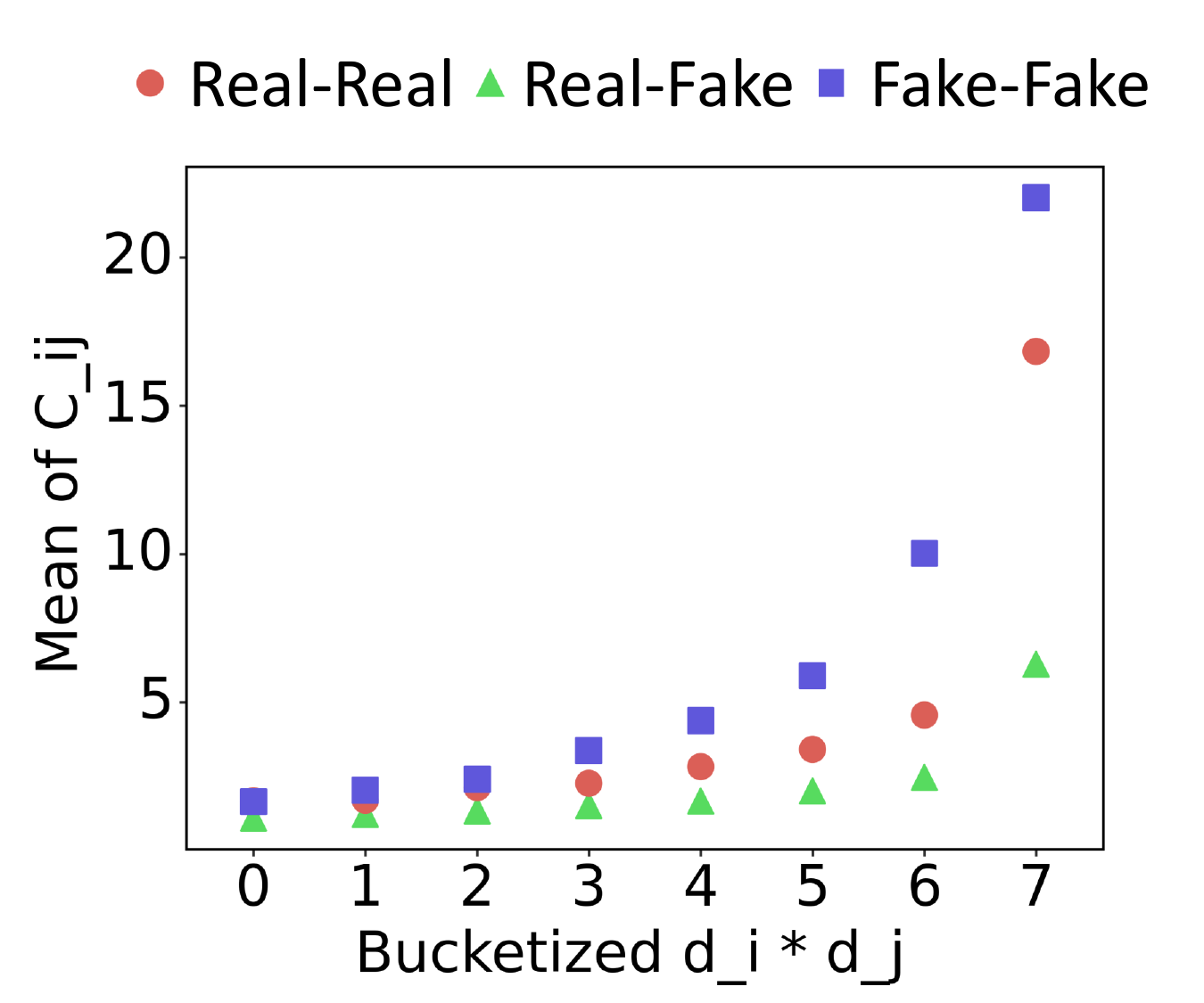}}
    \subfigure[\textbf{GossipCop}]{\includegraphics[width=0.49\columnwidth]{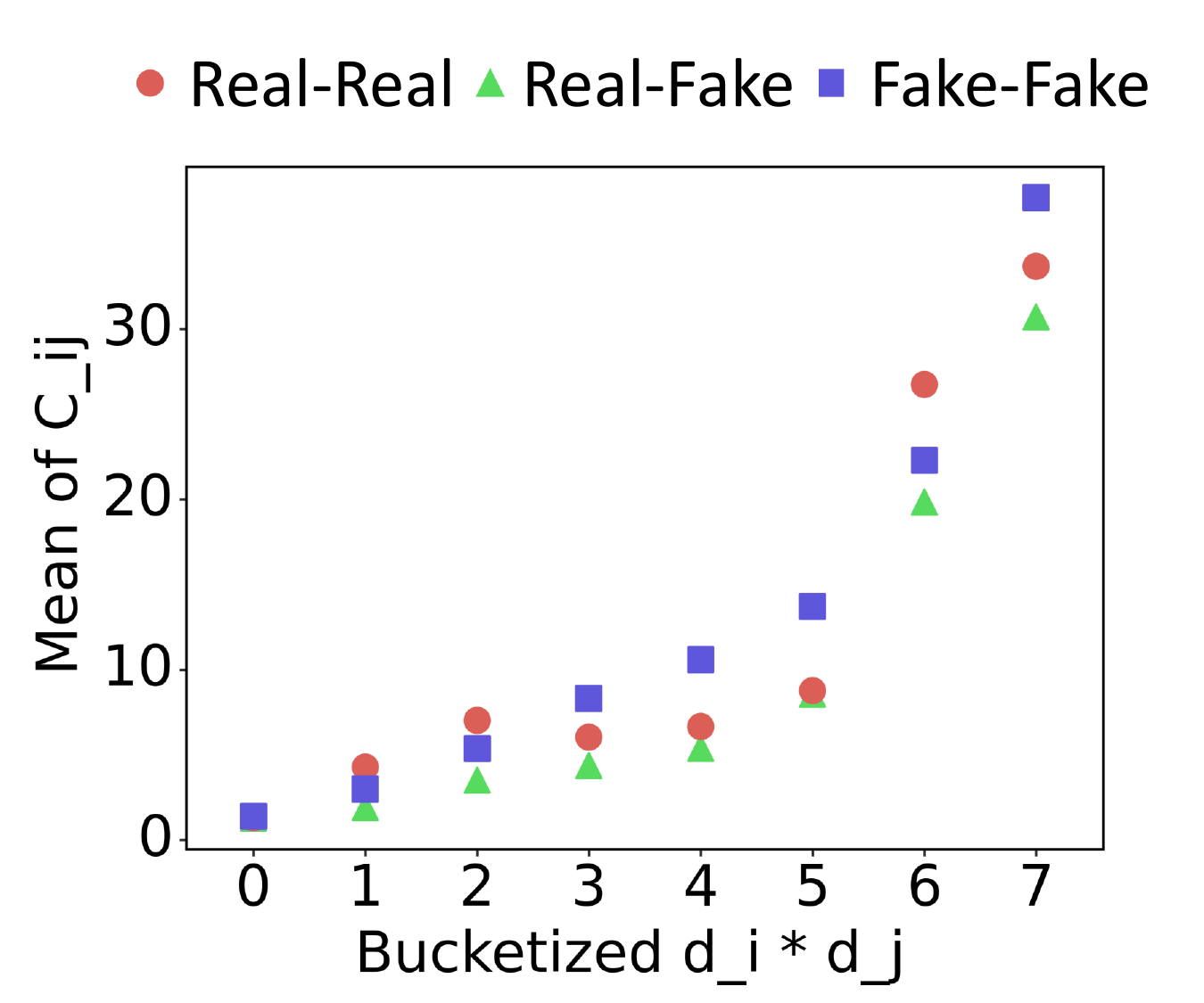}}
    \caption{News co-engagement patterns of news article pairs.  Edges in $\mathcal{G}$ represent shared readership between articles, and are grouped based on the articles' veracity labels.}
    \label{fig:co-engagement}
\end{figure}

\subsection{Empirical Observations}
\label{sec:empirical-obs}

In this subsection, we conduct preliminary analysis on real-world news to explore the veracity-related structural properties on the news engagement graph. We observe that fake and real news exhibit distinctive patterns in terms of weighted node \textbf{degrees}, motivated by which we design a degree-based social graph refinement framework to mitigate the edge noise issue 
 in Section \ref{sec:decor}. Our analysis is based on the FakeNewsNet \cite{shu2018fakenewsnet} benchmark, which consists of the PolitiFact and GossipCop datasets.

\subsubsection{Degree-Veracity Correlations.}

We first explore how the degree of a news article node is related to its veracity label. In other words, \textbf{do fake news articles attract more or less user engagements than real news}? 

Recall that we have a news engagement graph $\mathcal{G}=\{\mathcal{P},\mathcal{E}\}$ with adjacency matrix $\mathbf{A}$. The weighted node degrees in $\mathbf{A}$ can be used to measure the intensity of user engagements for each news article. In Figure \ref{fig:degree-cred}, we visualize the degree distributions of fake and real news with a kernel distribution estimation (KDE) plot, which depicts the node degrees with a continuous probability density curve. We make the following observation:

\begin{obs}\label{obs_deg}
On the news engagement graph, the degree distributions of nodes representing fake and real news articles show a clear difference. Note that different datasets can exhibit varying domain-specific patterns; for instance, in the GossipCop dataset containing celebrity news, real news tend to attract more engagements from active social users. However, this pattern does not apply to the politics-related PolitiFact dataset.
\end{obs}

\subsubsection{News Co-Engagement.} 
Next, we explore the degree-related properties of news article pairs connected by common readers (i.e. active social users in $\mathcal{U}_A$). Intuitively, given a pair of news articles $p_i$ and $p_j$ that share at least 1 reader, the corresponding edge $e_{ij}\in\mathcal{E}$ in the news engagement graph can be divided into three groups according to the veracity labels of $p_i$ and $p_j$: (1) real news pairs; (2) real-fake pairs; and (3) fake news pairs.

To quantify the shared user engagements between news article nodes w.r.t. the corresponding degrees, we compute a \textit{``co-engagement''} score $C_{ij}$ for news articles $p_i$ and $p_j$, formulated as: 
\begin{defn}[News Co-Engagement] 
$$
C_{ij} = |\mathcal{U}_i\cap\mathcal{U}_j|,
$$
where $\mathcal{U}_i\subset\mathcal{U}$ and $\mathcal{U}_j\subset\mathcal{U}$ are the sets of social users that engage with $p_i$ and $p_j$, respectively.
\end{defn}
We investigate the following question: \textbf{given an edge, are there any associations between its group, and the news co-engagement of the two nodes it connects?} In Figure \ref{fig:co-engagement}, we bucketize the edges by the value of $d_i\times d_j$, and plot the news co-engagement scores w.r.t. the edge groups. Note that here we adopt the product of degrees to distinguish edges with high values for both $d_i$ and $d_j$, and also motivated by our theoretical results in Section \ref{sec:dcsbm}.
Across the buckets, we observe the following pattern on news co-engagement:

\begin{obs} \label{obs_cij}
Given the degrees, fake news pairs tend to have higher $C_{ij}$ (i.e. more common users than expected given the degrees), while real-fake pairs tend to have lower $C_{ij}$ than both real news pairs and fake news pairs.
\end{obs}

Our two empirical observations provide distinctive degree-related cues pertaining to nodes (i.e. news articles) and edges (i.e. user engagements) on the news engagement graph (extended analysis and discussion are relegated to Appendix \ref{sec:extend-obs}). These patterns can guide a model in suppressing the noisy edges, as they can be leveraged to identify which edges are more likely to connect news articles of the same veracity. Meanwhile, we find that differences in the degree distributions can be complex (e.g., as shown in Figure \ref{fig:degree-cred}, fake news attract more user engagements than real news in PolitiFact, but less in GossipCop). This motivates our following degree-based innovations for a learnable social graph refinement approach. 

\begin{figure}[t]
    \centering
    \includegraphics[width=\columnwidth]{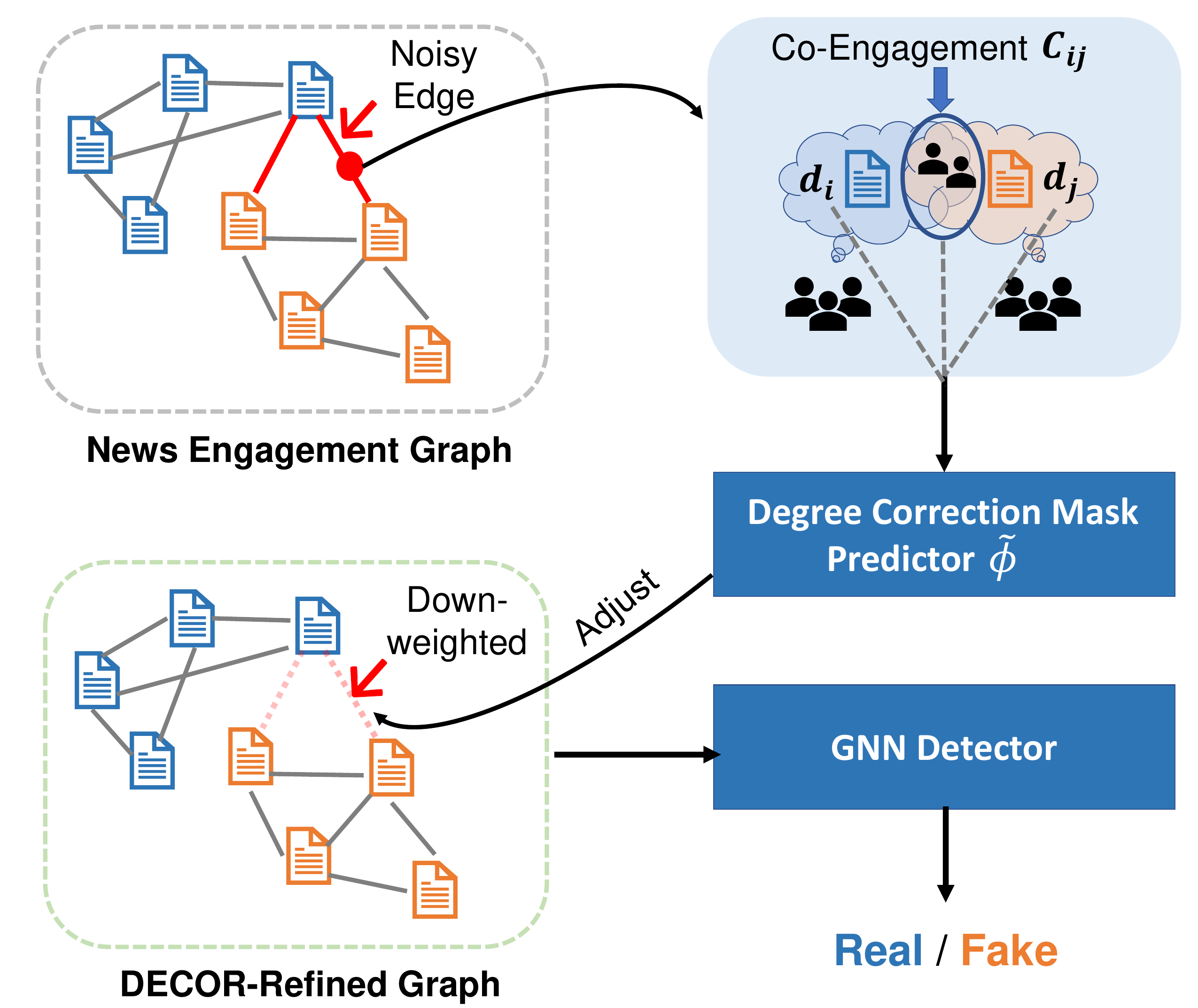}
    \caption{Overview of the proposed Degree-Corrected Social Graph Refinement (DECOR) framework.}
    \label{fig:framework}
\end{figure}

\section{Proposed Framework -- DECOR}
\label{sec:decor}

Motivated by our empirical findings on veracity-related degree patterns, we propose the DECOR framework for degree-corrected social graph refinement (overviewed in Figure \ref{fig:framework}). DECOR can be considered as a novel extension of the Degree-Corrected Stochastic Blockmodel (DCSBM) \cite{karrer2011stochastic} to the fake news detection scenario, which empowers fake news detectors with effective denoising of user engagements. Given a pair of news articles connected by common users, we propose a social degree correction module to adjust the corresponding edge weight using degrees and the news co-engagement. This module is jointly optimized with the GNN classifier, which leverages the corrected edge weights and news article features to predict the news veracity labels.

\subsection{Connection with the DCSBM Model}\label{sec:dcsbm}

In Section \ref{sec:empirical-obs}, we observed that degree patterns are closely related to news veracity labels.
Next, we formally demonstrate these connections from a theoretical perspective based on the DCSBM model \cite{karrer2011stochastic}, a generative model for graphs that derives edge placement likelihoods in a degree-based manner. The benefit of DCSBM is that it allows us to simultaneously model the effect of \emph{degree patterns} and \emph{class labels}, which are of key interest, in a tractable probabilistic way. Based on the DCSBM model, we will then theoretically derive a principled likelihood ratio-based approach for graph structure learning for the fake news detection application.

\paragraph{Framework} We first formulate the standard DCSBM under our fake news detection scenario. Recall the news engagement graph $\mathcal{G}=\{\mathcal{P},\mathcal{E}\}$ formulated in Section \ref{sec:newsgraph}, where $|\mathcal{P}|=N$. Each news article node in $\mathcal{G}$ is associated with a class label from the label space $\mathcal{Z}=\{0,1\}$. Consider a pair of news article nodes $p_i\in\mathcal{P}$ and $p_j\in\mathcal{P}$ with co-engagement $C_{ij}$. The nodes have class labels $z_i\in\mathcal{Z}$ and $z_j\in\mathcal{Z}$, respectively. Recall that $C_{ij}$ is defined as the number of common users between $p_i$ and $p_j$. 

Next, to formulate structure learning under the DCSBM model, our basic intuition is that \emph{same-class edges} (i.e., edges $e_{ij}$ where $z_i = z_j$) are more likely to be useful and informative than \emph{cross-class edges} (i.e., edges where $z_i \neq z_j$), and hence, structure learning should aim to give a higher weight to same-class edges. Intuitively, cross-class edges tend to indicate noisy edges, as in the example in Figure \ref{fig:noise-example}, where the co-engagement between them arises just by chance. Moreover, since our main goal is to classify $p_i$, identifying edges where $z_i = z_j$ clearly provides highly useful information for this task. Hence, our key idea is to perform structure learning by deriving the \emph{same-class likelihood ratio}:
\begin{defn}[Same-class likelihood ratio]
The same-class likelihood ratio, i.e. the likelihood ratio for $z_i = z_j$ over $z_i \neq z_j$ when observing $C_{ij}$ edges between $p_i$ and $p_j$, is  
\begin{align}
    LR_{ij} := \frac{\mathbb{P}(C_{ij}|z_i=z_j)}{\mathbb{P}(C_{ij}|z_i\neq z_j)}.
\end{align}
\end{defn}
The higher this likelihood ratio, the more evidence the data (specifically, $C_{ij}$) gives in favor of $z_i = z_j$ over $z_i \neq z_j$; and hence, structure learning should give a higher weight to such edges.

\paragraph{Derivation} Under the DCSBM model, the $C_{ij}$ edges between $p_i$ and $p_j$ are independently Poisson distributed, i.e., $C_{ij}\sim\mathsf{Poi}(\lambda_{ij})$, where $\lambda_{ij}$ denotes the expected number of edges:
\begin{equation}
    \lambda_{ij}= \begin{cases} 
        \beta_i \beta_j p & \text { if } z_i = z_j \\ 
        \beta_i \beta_j q & \text { if } z_i \neq z_j 
    \end{cases},
\label{eq:lambda_ij}
\end{equation}
where $\beta_i$ and $\beta_j$ are the ``degree correction parameters'' that allow us to generate nodes with different degrees. $p$ and $q$ are parameters controlling the rate at which edges are generated under the same-class and cross-class cases, respectively. Generally, we have $p>q$, i.e., same-class edges have a higher tendency to be generated.

The corresponding maximum likelihood values $\hat{\beta{_i}}$ and $\hat{\beta{_j}}$ for $\beta_i$ and $\beta_j$ are given as
\begin{equation}
    \hat{\beta{_i}} = \frac{d_i}{m},\quad \hat{\beta{_j}} = \frac{d_j}{m},
\label{eq:mle}
\end{equation}
in the DCSBM model \cite{karrer2011stochastic}, where $m=|\mathcal{E}|$ denotes the number of edges. $d_i$ and $d_j$ respectively refer to the weighted degrees of nodes $p_i$ and $p_j$.

Since $C_{ij}\sim\mathsf{Poi}(\lambda_{ij})$, the likelihood ratio $LR_{ij}$ for $z_i=z_j$ over $z_i\neq z_j$ can be derived as:
\begin{equation} 
\begin{split}
    LR_{ij} &= \frac{\mathbb{P}(C_{ij}|z_i=z_j)}{\mathbb{P}(C_{ij}|z_i\neq z_j)} \\
    & = \frac{e^{-\beta_i \beta_j p}(\beta_i \beta_j p)^{C_{ij}}}{e^{-\beta_i \beta_j q}(\beta_i \beta_j q)^{C_{ij}}} \\
    & = e^{-\beta_i\beta_j(p-q)}(\frac{p}{q})^{C_{ij}}.
\end{split}
\label{eq:likelihood-ratio}
\end{equation}
Substituting the $\hat{\beta{_i}}$ and $\hat{\beta{_j}}$ given in Eq.\ref{eq:mle} into Eq.\ref{eq:likelihood-ratio}, we derive the maximum likelihood estimate for $LR_{ij}$:
\begin{equation} 
   \mathsf{MLE}(LR_{ij})=e^{-\frac{d_i d_j}{m^2}(p-q)}(\frac{p}{q})^{C_{ij}}.
\label{eq:sub-mle}
\end{equation}
Treating $m, p, q$ as fixed (since they are shared by all nodes), we thus see that the MLE is a function of $C_{ij}$, $d_i$ and $d_j$: in particular, it is a \emph{log-linear} function of $C_{ij}$ and $d_i d_j$:
\begin{align} 
   \mathsf{MLE}(LR_{ij})&={\Phi}(C_{ij}, d_i, d_j) := e^{-\frac{d_i d_j}{m^2}(p-q)}(\frac{p}{q})^{C_{ij}} \\ 
   &= \exp\left[ \begin{pmatrix} C_{ij} \\ d_{i} d_j \end{pmatrix} \cdot \begin{pmatrix} \log(p)-\log(q) \\ -\frac{p-q}{m^2} \end{pmatrix} \right] \label{eq:Phi}
\end{align}

\paragraph{Implications} We first note that Eq. \ref{eq:Phi} agrees with our empirical finding in Observation \ref{obs_cij}: if we fix $d_i d_j$ in Eq. \ref{eq:Phi}, then as long as $\log(p)-\log(q) > 0$, we observe that higher $C_{ij}$ is associated with a higher $LR_{ij}$, and thus a higher probability of same-class edges ($z_i = z_j$), agreeing with Figure \ref{fig:co-engagement} where the Real-Fake edges have lowest $C_{ij}$ for a given $d_i d_j$.

For structure learning purposes, we could simply use ${\Phi}(C_{ij}, d_i, d_j)$, which we recall is an estimator for $LR_{ij} = \frac{\mathbb{P}(C_{ij}|z_i=z_j)}{\mathbb{P}(C_{ij}|z_i\neq z_j)}$. However, the standard DCSBM model is built upon relatively strong assumptions (e.g. pre-defined $p$ and $q$ values); for fitting real data, we would like to relax these assumptions and allow the model to be flexibly learned from data. The DCSBM model contains very few learnable parameters, which is a fundamental limitation in adapting to the complex degree-based patterns in the news engagement graph. This motivates us to develop DECOR, a degree-based learnable social graph refinement framework, which we will next describe in detail, by \emph{relaxing the assumption of log-linearity}: that is, instead of treating ${\Phi}(C_{ij}, d_i, d_j)$ as a fixed and log-linear function defined in Eq. \ref{eq:Phi}, we instead treat it as a \emph{learnable non-linear function $\Tilde{\Phi}(C_{ij}, d_i, d_j)$} to be updated jointly with the rest of the model, during the structure learning process. 

\subsection{Social Degree Correction}

As illustrated in Figure \ref{fig:noise-example}, the news engagement graph contains structural noise. In light of our empirical findings on degree-veracity relationships and the DCSBM framework, we propose to learn a degree-corrected social graph that downweights the noisy edges to eliminate their negative impacts and facilitate fake news detection via GNN-based classifiers.

Recall that the type of an edge in the news engagement graph (i.e. connecting new articles of same or different veracity) is characterized by the co-engagement and degrees of the connected articles. Motivated by the DCSBM model's degree-based probabilistic derivation of edge placement likelihood, we propose to adjust edge weights in the news engagement graph via learning a \textit{social degree correction mask} $\mathbf{M}\in\mathbb{R}^{N\times N}$, where $\mathbf{M}_{ij}$ in the interval $(0,1)$ represents the degree correction score for edge $e_{ij}$ between news article nodes $p_i$ and $p_j$.

The value of $\mathbf{M}_{ij}$ is predicted given co-engagement $C_{ij}$ of articles $p_i$ and $p_j$, and the articles' weighted node degrees $d_i$ and $d_j$ from the news engagement graph. Specifically, we adopt a neural predictor to obtain $\mathbf{s}_{ij}\in\mathbb{R}^2$, which contains two scores for edge preservation and elimination, respectively:
\begin{equation}
    \mathbf{s}_{ij}=\Tilde{\Phi}(C_{ij}, d_i,d_j).
\label{eq:edgepred}
\end{equation}
$\Tilde{\Phi}(\cdot)$ is a MLP-based architecture, and can be considered as a learnable extension of Eq.\ref{eq:Phi} in the DCSBM model.  

The scores in $\mathbf{s}_{ij}$ are normalized via the softmax function. To preserve computational efficiency, we design the social degree correction process as \textit{pruning}. In other words, we conduct Eq.\ref{eq:edgepred} on all the news pairs connected by common users to obtain the corresponding degree correction scores: 
\begin{equation}
    \mathbf{M}_{i j}= \begin{cases}v_{ij} & \text { if } C_{ij}\neq 0\\ 
    0 & \text { else }\end{cases}.
\label{eq:mask}
\end{equation}
where $v_{ij}$ denotes the softmax-normalized score in $\mathbf{s}_{ij}$  that correlates with edge preservation. 

Given the co-engagement matrix $\mathbf{C}$ of news engagement graph $\mathcal{G}$, we utilize $\mathbf{M}$ to obtain a degree-corrected adjacency matrix $\mathbf{A}_c$:
\begin{equation}
    \mathbf{\hat{A}} = \mathbf{C}\cdot \mathbf{M} + \mathbf{I}
\label{eq:corrected1}
\end{equation}
\begin{equation}
    \mathbf{A}_{c} = \mathbf{D}^{-\frac{1}{2}}
    \mathbf{\hat{A}} \mathbf{D}^{-\frac{1}{2}} ,
\label{eq:corrected2}
\end{equation}
where $\mathbf{I}$ represents an identity matrix of size $N$, and $\mathbf{D}$ is the diagonal matrix of degrees for $\mathbf{\hat{A}}$.

Through the above operations, noisy edges in the news engagement graph are assigned smaller weights, as $\Tilde\Phi(\cdot)$ in Eq.\ref{eq:edgepred} leverages degree-based properties to predict a low degree correction score.

\subsection{Prediction on Degree-Corrected Graph}

With the degree-corrected adjacency matrix $\mathbf{A}_{c}$, we can leverage the powerful GNN architectures (e.g. GCN \cite{kipf2017semi}, GIN \cite{xu2018how} and GraphConv \cite{morris2019graphconv}) to predict the veracity labels of article nodes in the degree-corrected news engagement graph. 

Central to GNNs is the message-passing mechanism \cite{gilmer2017neural}, which follows an iterative scheme of updating node representations based on information aggregation among the node's neighborhood. For a news article $p\in\mathcal{P}$, the initial news article feature $\mathbf{h}_p^{(0)}$ is set as the news content representation $\mathbf{x}_p$:
\begin{equation}
    \mathbf{h}_p^{(0)} = \mathbf{x}_p,
\label{eq:feature-init}
\end{equation}
where $\mathbf{x}_p$ is extracted from news article $p$ via a pre-trained language model $\mathcal{M}$ with frozen parameters. At the $k$-th layer of a GNN, the news article representation $\mathbf{h}_p^{(k)}$ is obtained via:   
\begin{equation}
 \mathbf{m}_p^{(k)}=\mathsf{AGGREGATE}^{(k)}\left(\left\{\mathbf{h}_u^{(k-1)}, \forall u \in \mathcal{N}(p)\right\}\right)
\label{eq:gnn1}
\end{equation}
\begin{equation}
\mathbf{h}_p^{(k)}=\mathsf{COMBINE}^{(k)}\left(\mathbf{h}_p^{(k-1)}, \mathbf{m}_p^{(k)}\right),
\label{eq:gnn2}
\end{equation} 
where $\mathcal{N}(p)$ denotes the neighbors of $p$ on the news engagement graph, and $\mathbf{m}_p^{(k)}$ is the aggregated information from $\mathcal{N}(p)$.

Let $\mathbf{h}_p\in\mathbb{R}^2$ be the output of the GNN-based classifier for node $p$. Then, the news veracity label of $p$ is predicted as $\tilde{\mathbf{y}}_p=\mathsf{softmax}(\mathbf{h}_p)$. During training, we minimize the following cross entropy loss:
\begin{equation}
    \mathcal{L}=\sum_{p \in \mathcal{P}_{train}} \mathsf{CELoss}\left(\tilde{\mathbf{y}}_p, \mathbf{y}_p\right).
\label{eq:loss}
\end{equation}

The degree correction mask predictor $\Tilde{\Phi}(\cdot)$ is jointly optimized with the GNN-based classifier. DECOR utilizes low-dimensional degree-related properties to guide the social degree correction operations, which facilitates edge denoising on $\mathcal{G}$ without loss of computational efficiency.

\begin{table}[t]

\caption{Dataset statistics.}
{
 \begin{tabular}{lcc} \toprule
 \textbf{Dataset} &  \textbf{PolitiFact} & \textbf{GossipCop} \\ 
 \toprule
 \# News Articles & 497 & 16,599 \\
 \# Real News & 225 & 12,641 \\
 \# Fake News & 272 & 3,958 \\  
 \# User-News Engagements & 227,184 & 963,009 \\ 
 \# Distinct Users & 143,481 & 202,907 \\ 
 \bottomrule
\end{tabular} 
}
 \label{tab:ds-stats}
\end{table}

\begin{table*}[ht]
\caption{Performance comparison between DECOR and baselines (G1: Content-based, G2: Graph-based, and G3: GSL-based). Bold and underline indicates the best overall and baseline performance, respectively. $^{*}$ denotes that DECOR performs significantly better than the corresponding GNN backbone at $p<0.01$ level using the Wilcoxon signed-rank test.}
\centering 
{
 \begin{tabular}{clcccccccccc} \toprule
 \multirow{2}{*}{}& \multirow{2}{*}{\textbf{Method}} & \multicolumn{4}{c}{\textbf{PolitiFact}} && \multicolumn{4}{c}{\textbf{GossipCop}} \\
 \cmidrule{3-6} \cmidrule{8-11} 
 && Acc. & Prec. & Rec. & F1 &&  Acc. & Prec. & Rec. & F1 \\ 
 \toprule
  \multirow{5}{*} {\textbf{G1}}& dEFEND\textbackslash c \cite{shu2019defend} & 0.8207 & 0.8226 & 0.8198 & 0.8195 && 0.7998 & 0.7251 & 0.6787 & 0.6936 \\
 &SAFE\textbackslash v \cite{zhou2020safe} & 0.8146 & 0.8185 & 0.8114 & 0.8121 && 0.7920 & 0.7090 & 0.6739 & 0.6866 \\
 &SentGCN \cite{vaibhav2019sentence}  & 0.8404 & 0.8567 & 0.8446 & 0.8385 && 0.7935 & 0.7130 & 0.6797 & 0.6913 \\
 &BERT \cite{devlin2019bert} & 0.8586 & 0.8692 & 0.8584 & 0.8564 &&  0.8498 & 0.8084 & 0.7480 & 0.7703 \\
 &DistilBERT \cite{sanh2019distil} & 0.8419 & 0.8518 & 0.8426 & 0.8403 &&  0.8349 & 0.7765 & 0.7422 & 0.7563 \\
 \midrule
\multirow{5}{*} {\textbf{G2}}& GCNFN \cite{monti2019fake} & 0.8687 & 0.8694 & 0.8680 & 0.8674 && 0.8520 & 0.9016 & 0.6950 & 0.7354 \\
  &FANG \cite{nguyen2020fang}  & 0.8601 & 0.8625 & 0.8647 & 0.8599 &&  0.8715 & 0.8835 & 0.7506 & 0.7897 \\
  &GCN \cite{kipf2017semi} & 0.8965 & 0.9027 & 0.9034 & 0.8964 && 0.9146 & \underline{0.9370} & 0.8281 & 0.8668 \\
  &GIN \cite{xu2018how} & \underline{0.9025} & \underline{0.9037} & \underline{0.9068} & \underline{0.9024} && \underline{0.9197} & 0.9308 & \underline{0.8450} & \underline{0.8780} \\
  &GraphConv \cite{morris2019graphconv} & 0.8990 & 0.9007 & 0.9037 & 0.8989 && 0.9150 & 0.9172 & 0.8426 & 0.8720 \\
 \midrule
 
  \multirow{2}{*} {\textbf{G3}}&Pro-GNN \cite{jin2020graph} & 0.7747 & 0.7779 & 0.7690 & 0.7691 && \multicolumn{4}{c}{\multirow{2}{*}{OOM}}  \\
 &RS-GNN \cite{dai2022towards} & 0.7439 & 0.7839 & 0.7253 & 0.7210 && \multicolumn{4}{c}{} \\
 \midrule
  \multirow{3}{*} {\textbf{Ours}}&GCN w/ DECOR & \textbf{0.9480}$^{*}$ & \textbf{0.9483}$^{*}$ & \textbf{0.9520}$^{*}$ & \textbf{0.9479}$^{*}$ && 0.9310$^{*}$ & \textbf{0.9377} & 0.8694$^{*}$ & 0.8973$^{*}$ \\
  & GIN w/ DECOR  & 0.9399$^{*}$ & 0.9398$^{*}$ & 0.9433$^{*}$ & 0.9398$^{*}$ && \textbf{0.9333}$^{*}$ & 0.9294 & \textbf{0.8828}$^{*}$ & \textbf{0.9031}$^{*}$ \\
  & GraphConv w/ DECOR  & 0.9379$^{*}$ & 0.9376$^{*}$ & 0.9412$^{*}$ & 0.9377$^{*}$ && 0.9259$^{*}$ & 0.9130 & 0.8805$^{*}$ & 0.8945$^{*}$ \\
 \bottomrule
\end{tabular} 
}
\label{tab:performance}
\end{table*}

\section{Experiments}

In this section, we empirically evaluate DECOR to answer the following five research questions:
\begin{itemize}[leftmargin=*]
    \item \textbf{Fake News Detection Performance} (Section \ref{sec:perfomance}):  How well does DECOR perform compared with competitive baselines?
    \item \textbf{Ablation Study} (Section \ref{sec:ablation}): How effective are co-engagement and degree patterns, respectively, in improving the fake news detection performance of DECOR?  
    \item \textbf{Limited Training Data} (Section \ref{sec:limited-data}): Does DECOR perform well under label sparsity?
    \item \textbf{Computational Efficiency} (Section \ref{sec:efficiency}): How efficient is DECOR compared with existing GSL methods?
    \item \textbf{Case Study} (Section \ref{sec:case-study}): Does DECOR downweight the noisy edges connecting influential real and fake news articles?

\end{itemize}

\subsection{Experimental Setup}
\label{sec:setup}

\subsubsection{Datasets}
We evaluate DECOR on the public benchmark FakeNewsNet \cite{shu2018fakenewsnet}, which consists of two real-world datasets: PolitiFact and GossipCop. Both datasets contain news articles annotated by leading fact-checking websites and the articles' related social user engagements from Twitter.
The descriptive statistics of the datasets are summarized in Table \ref{tab:ds-stats}. 

To simulate the real-world scenarios, we split the news samples following a \textit{temporal} order. Specifically, the most recent $20$\% real and fake news instances constitute the test set, and the remaining $80$\% instances posted earlier serve as the training set.

\subsubsection{Baselines} 

We benchmark DECOR against twelve representative baseline methods, which can be categorized into the following three groups by model architecture:

\textbf{News content based methods (G1)} leverage the semantic features in the news articles. Specifically, \textbf{dEFEND\textbackslash c} is a content-based variant of dEFEND \cite{shu2019defend} without incorporating user comment texts, which utilizes a hierarchical network with the co-attention mechanism. \textbf{SAFE\textbackslash v} is a content-based variant of SAFE \cite{zhou2020safe} without incorporating visual information from images, which leverages a CNN-based fake news detector. \textbf{SentGCN} \cite{vaibhav2019sentence} models each news article as a graph of sentences, and utilize the GCN \cite{kipf2017semi} architecture for news veracity prediction. \textbf{BERT} \cite{devlin2019bert} and \textbf{DistilBERT} \cite{sanh2019distil} (with model names BERT-base and DistilBERT-base, respectively) are large pre-trained bidirectional Transformers, which we fine-tune to the downstream task of fake news detection. 

\textbf{Social graph based methods (G2)} encode the social context into graph structures, and leverage GNNs to learn news article representations. Specifically, \textbf{GCNFN} \cite{monti2019fake} leverages user responses and user following relations to construct a propagation tree for each news article. \textbf{FANG} \cite{nguyen2020fang} establishes a comprehensive social graph with users, news and sources, and learns the representations with GraphSAGE \cite{hamilton2017inductive}.  We also apply three representative GNN architectures on our proposed news engagement graph, namely \textbf{GCN} \cite{kipf2017semi}, \textbf{GIN} \cite{xu2018how}, and \textbf{GraphConv} \cite{morris2019graphconv}. For a fair comparison, we only implement the model components involving news articles, social user identities, and user-news relations.

\textbf{Graph Structure learning (GSL) methods (G3)} aim to enhance representation via learning an optimized graph structure. We implement two GSL methods that focus on edge denoising, \textbf{Pro-GNN} \cite{jin2020graph} applies low-rank and sparsity properties to learn a clean graph structure that is similar to the original graph. \textbf{RS-GNN} \cite{dai2022towards} simultaneously learns a denoised graph and a robust GNN via constructing a link predictor guided by node feature similarity.

\subsubsection{Evaluation Metrics} 
Following prior works \cite{shu2019defend,zhou2020safe}, we adopt four widely-used metrics to evaluate the performance of fake news detection methods: Accuracy (\textbf{Acc.}), Precision (\textbf{Prec.}), Recall (\textbf{Rec.}) and F1 Score (\textbf{F1}). In all experiments, we report the average metrics across 20 different runs of each method.

\subsubsection{Implementation Details} We implement our proposed DECOR model and its variants based on PyTorch 1.10.0 with CUDA 11.1, and train them on a server running Ubuntu 18.04 with NVIDIA RTX 3090 GPU and Intel(R) Xeon(R) Gold 6226R CPU @ 2.90GHz. To construct DECOR's news engagement graph, we select active social users with least $3$ reposts, and threshold a user's maximum number of interactions with each news article at $1\%$ of the total number of news articles. We extract $768$-dimensional news article features via a pre-trained BERT model with frozen parameters; specifically, we utilize pre-trained weights from HuggingFace Transformers 4.13.0 \cite{wolf20transformers} (model name: bert-base-uncased). The predictor $\Tilde{\Phi}(\cdot)$ for social degree correction is a $2$-layer MLP with hidden size $16$ for PolitiFact and $8$ for GossipCop. The GNN architecture is set to $2$ layers with $64$ hidden dimensions. The model is trained for $800$ epochs, and model parameters are updated for via an Adam optimizer \cite{kingma2014adam} with learning rate $0.0005$. 

Technically, our framework is model-agnostic, which could coordinate with various GNN models on the news engagement graph. Here, we select three representative GNN architectures as backbones: GCN \cite{kipf2017semi}, GIN \cite{xu2018how} and GraphConv \cite{morris2019graphconv}. For the implementation of baseline methods, we follow the architectures and hyperparameter values suggested by their respective authors.

\subsection{Performance Comparison}
\label{sec:perfomance}
This subsection compares DECOR with various content-based, graph-based and GSL baselines on fake news detection.

Table \ref{tab:performance} shows that DECOR consistently outperforms the competitive baseline methods by significant margins ($p<0.01$). We make the following observations:
 \textit{\textbf{(1)}} Among the five content-based methods (G1), pre-trained language models (PLMs) outperform the ``train-from-scratch'' methods. The effectiveness of PLMs demonstrates the benefits of pre-training on large-scale corpora, from which the model obtains rich semantic knowledge. \textit{\textbf{(2)}} Methods that incorporate social graphs (G2) consistently outperforms the content-based methods (G1). This signifies the importance of user engagement patterns, and indicates that exploiting social context is central to effective fake news detection. \textit{\textbf{(3)}} Among the social graph based methods (G2), methods that leverage our proposed news engagement graph (GCN, GIN, GraphConv) outperform methods that formulate both news and users as graph nodes (GCNFN and FANG). Our graph formulation is superior, in that it focuses solely on the co-engagement of social users; it facilitates direct information propagation among articles with similar reader groups, and avoids the potential task-irrelevant signals from user profiles and related tweets. \textit{\textbf{(4)}} Existing GSL methods for edge denoising (G3) are not suited to fake news detection. One possible reason is that these methods are similarity-guided, i.e., links between nodes of dissimilar features are strongly suppressed. However, in our fake news detection scenario, two news articles on different topics can be closely connected in terms of co-engagement and veracity type. \textit{\textbf{(5)}} Compared with competitive fake news detectors, DECOR substantially enhances the performance of three representative GNN backbones. This validates the effectiveness of using degrees and co-engagement to learn a refined news engagement graph. 

\begin{figure}[t]
    \centering
    \includegraphics[width=\columnwidth]{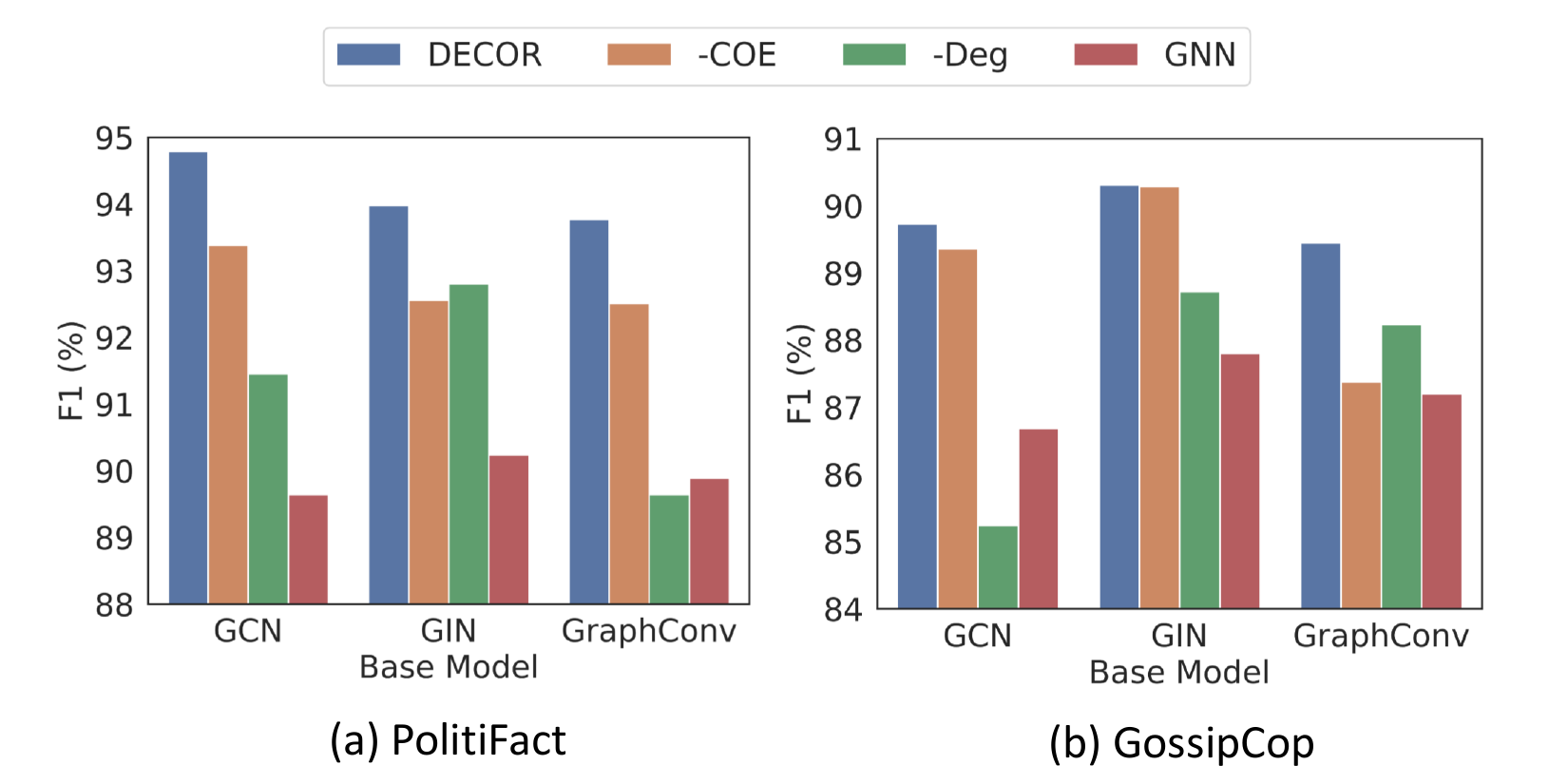}
    \caption{Ablation study of DECOR.}
    \label{fig:ablation}
\end{figure}

\subsection{Ablation Study}
\label{sec:ablation}

We conduct an ablation study to assess the contribution of DECOR's major components in detecting fake news, and summarize the results in Figure \ref{fig:ablation}. We compare DECOR with two variants, namely \textbf{DECOR-COE} without co-engagement, and \textbf{DECOR-Deg} without degrees (definitions given in Section \ref{sec:empirical-obs}).

As shown in Figure \ref{fig:ablation}, comparing DECOR with either DECOR-COE or DECOR-Deg, the superior fake news detection performance of DECOR illustrates that both co-engagement and degrees play a significant role in achieving the final improvements. Note that in numerous cases, DECOR-COE guided by degrees outperforms the corresponding GNN backbones that utilize the raw news engagement graph, which is consistent with our first empirical finding (Observation \ref{obs_deg}) on the distinctive connections between node degrees and news veracity. This further highlights the effectiveness of incorporating degree-related properties for social graph refinement.

\begin{table}[t]
\caption{Model efficiency comparison on PolitiFact dataset. }
{
 \begin{tabular}{lccc} \toprule
 Method &  \# Params & Runtime (s) & Avg. F1 \\ 
 \toprule
 GCN & 49,280 & 2.47 & 0.8964 \\
 GIN & 49,347 & 3.19 & 0.9024 \\
 GraphConv & 98,627 & 3.72 & 0.8989 \\  
 \midrule
 Pro-GNN & 296,289 & 158.03 & 0.7691 \\ 
 RS-GNN & 102,722 & 39.51 & 0.7210 \\ 
 \midrule
 GCN w/ DECOR & 49,410 & 4.63 & 0.9479 \\
 GIN w/ DECOR & 49,477 & 5.09 & 0.9398 \\
 GraphConv w/ DECOR & 98,757 & 5.17 & 0.9377 \\   
 
 \bottomrule
\end{tabular} 
}
 \label{tab:efficiency}
\end{table}

\begin{figure}[t]
    \centering
    \includegraphics[width=\columnwidth]{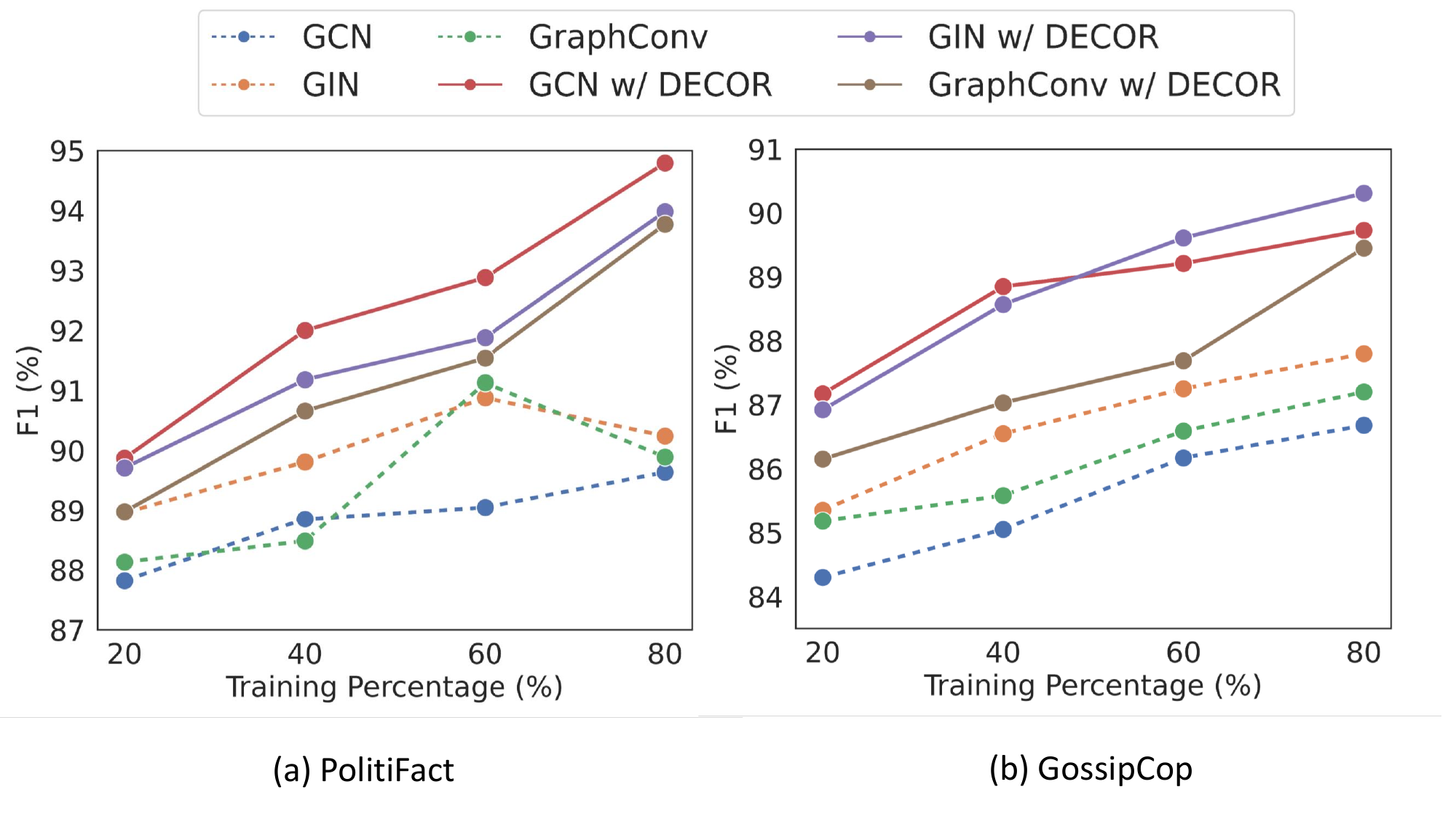}
    \caption{Comparison of DECOR against baselines (F1 Score) under varying training data sizes.}
    \label{fig:limited-data}
\end{figure}

\begin{figure*}[t]
    \centering
    \includegraphics[width=0.85\textwidth]{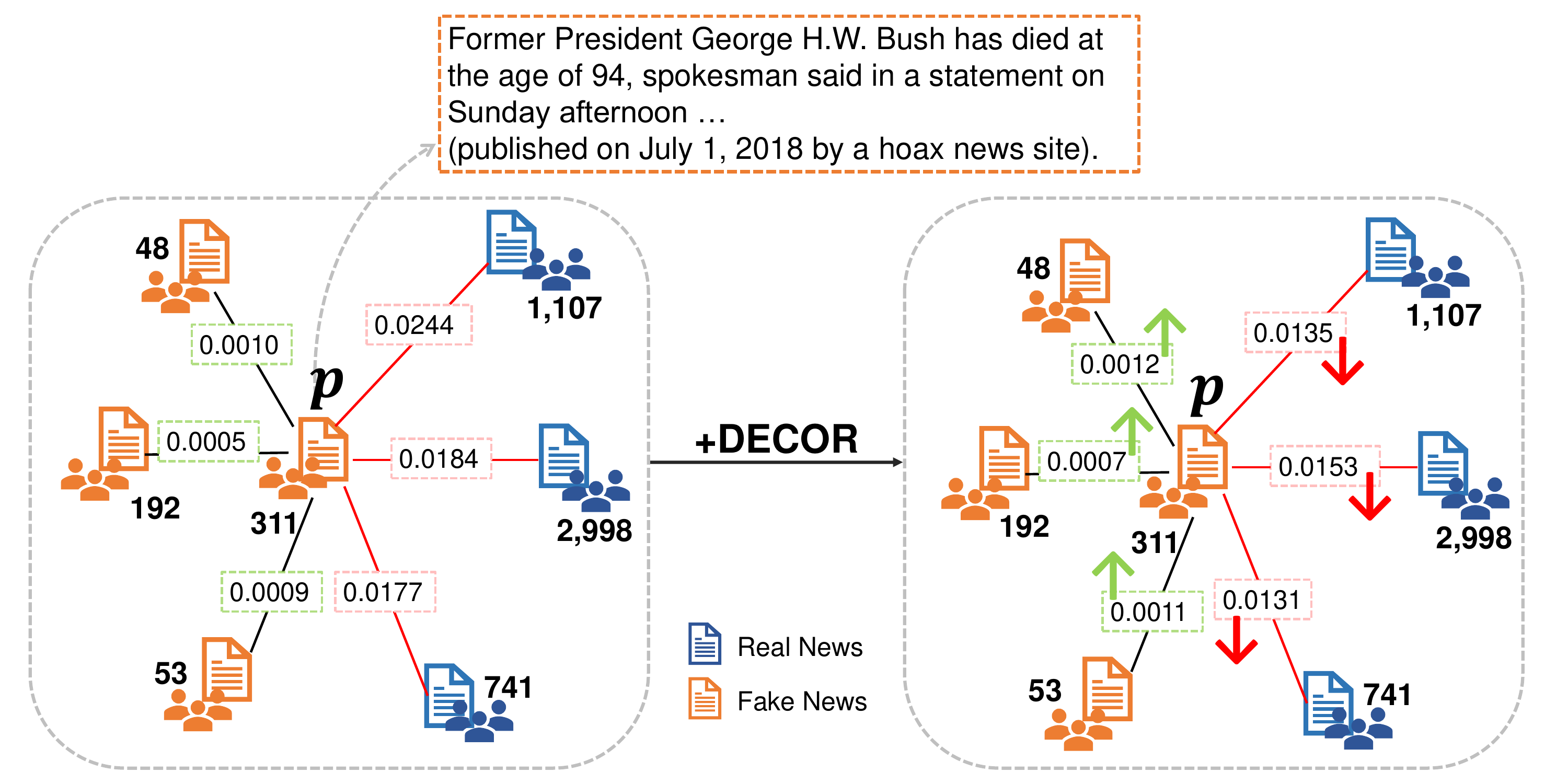}
    \caption{DECOR effectively downweights the noisy edges between influential real and fake news articles, while preserving the informative edges between news of the same veracity type. The edge weights are drawn from the normalized versions of the adjacency matrix $\mathbf{A}$ and the DECOR-refined $\mathbf{A}_c$, respectively. The number in bold font beside each user icon represents the number of user engagements associated with the corresponding news article. }
    \label{fig:case-study}
\end{figure*}

\subsection{Performance under Label Scarcity}
\label{sec:limited-data}

Label scarcity poses an imminent challenge for real-world applications of fake news detection. Due to the timely nature of news articles, high-quality annotations are usually scarce. We evaluate the performance of DECOR under limited training samples, and summarize the results in Figure \ref{fig:limited-data}. We observe that DECOR consistently outperforms the competitive GNN baselines on the news engagement graph for all training sizes: $20\%$, $40\%$, $60\%$ and $80\%$ of the data. DECOR learns an optimized graph by explicitly leveraging the degree-related structural signals embedded in degrees and news co-engagement, which serves as informative news veracity indicators and thereby complement the limited ground-truth knowledge from fact-checked annotations.

\subsection{Computational Efficiency}
\label{sec:efficiency}

We evaluate the computational cost of DECOR regarding parameter number and model runtime. Specifically, we train all models on the same GPU device for $800$ epochs, and compare the time elapsed. Note that both Pro-GNN and RS-GNN adopt the same 2-layer GCN architecture as the ``GCN'' method reported in Table \ref{tab:efficiency}.

Results in Table \ref{tab:efficiency} validate that DECOR is able to achieve impressive performance gains while maintaining low computational cost. Compared with existing GSL methods, three innovations account for DECOR's efficiency in fake news detection: \textit{\textbf{(1)}} DECOR leverages low-dimensional features (i.e. degrees and co-engagement) to predict an adjustment score for each edge, whereas existing GSL methods utilize node features that are high-dimensional in terms of news article representations. \textit{\textbf{(2)}} DECOR utilizes a lightweight degree correction component, which facilitates joint optimization of the social degree correction module and the GNN detector. In contrast, existing GSL methods adopt alternating optimization of the GNN and the link predictor, resulting in slower model training. \textit{\textbf{(3)}} DECOR operates as pruning on the existing edges in the news engagement graph, whereas existing GSL methods conduct pairwise computations (e.g. feature similarity) among all nodes. Hence, the complexity of DECOR is linear to the number of edges, whereas existing GSL methods incur up to quadratic complexity. These results suggest that DECOR is suitable for deployment in resource-limited scenarios, e.g., online fact-checking services.

\subsection{Case Study}
\label{sec:case-study}

To further illustrate why DECOR outperforms existing social graph based models and GSL methods, we conduct a case study to illustrate DECOR's capability of downweighting the noisy edges between fake and real news articles. 

In Figure \ref{fig:case-study}, we visualize exemplar cases in the neighborhood of $p$, an influential fake news article published by a hoax news site. From the subgraph on the left hand side, we observe that $p$ is involved in two types of edges: \textit{\textbf{(1)}} Noisy edges with large edge weights. $p$ is closely connected with three influential real news pieces. As these articles all focus on trending political topics, they attract a large number of common readers. \textit{\textbf{(2)}} Clean edges with small edge weights. $p$ is also connected with several fake news pieces; however, these articles attract less social users, which results in small groups of common readers with $p$. These structural patterns are problematic, as propagating information among noisy edges can contaminate the neighborhood, leading to suboptimal article representations. Existing social graph based models generally assume a fixed graph structure and are thereby heavily limited in suppressing edge noise. Prior works on similarity-guided edge denoising also cannot address this issue, as the articles contain similar topics but different veracity. In contrast, DECOR leverages the structural degree-based properties in a flexible manner. This facilitates the elimination of degree-related edge noise. From the subgraph on the right hand side of Figure \ref{fig:case-study}, we find that DECOR effectively suppresses the noisy edges, and recognizes the clean edges by assigning larger weights. These cases provide strong empirical evidence that DECOR effectively refines the news engagement graph for enhanced fake news detection.

\section{Conclusion and Future Work}

In this paper, we investigate the fake news detection problem from a novel aspect of social graph refinement. We observe that edge noise in the news engagement graph are degree-related, and find that news veracity labels closely correlate with two structural properties: degrees and news co-engagement. Motivated by the DCSBM model's degree-based probabilistic framework for  edge placement, we develop DECOR, a degree-based learnable social graph refinement framework. DECOR facilitates effective suppression of noisy edges through a learnable social degree correction mask, which predicts an adjustment score for each edge based on the aforementioned degree-related properties. Experiments on two real-world benchmarks demonstrate that DECOR can be easily plugged into various powerful GNN backbones as an enhancement. Furthermore, DECOR's structural corrections are guided by low-dimensional degree-related features, allowing for computationally efficient applications. We believe our empirical and theoretical findings will provide insights for future research in designing and refining more complex multi-relational social graphs for fake news detection.

\section{Acknowledgements}
This work was supported by NUS-NCS Joint Laboratory (A-0008542-00-00). The authors would like to thank the anonymous reviewers for their valuable feedback.

\bibliographystyle{ACM-Reference-Format}
\balance
\bibliography{reference}

\appendix
\begin{table*}[h!]
\caption{Descriptive statistics of news datasets with different topics.}
{
 \begin{tabular}{lcccccc} \toprule
 \textbf{Dataset} &  \textbf{PolitiFact} & \textbf{GossipCop} &  \textbf{FANG} & \textbf{MC-Fake (Syria War)}&  \textbf{MC-Fake (Health)} & \textbf{MC-Fake (Covid)}\\ 
 \toprule
 \# News Articles & 497 & 16,599 & 727 & 2,259 & 5,322 & 5,248 \\
 \# Real News & 225 & 12,641 & 419 & 2,082 & 4,788 & 4,524 \\
 \# Fake News & 272 & 3,958 & 308 & 177 & 534 & 724\\  
 \# User-News Engagements & 227,184 & 963,009  & 41,747 & 268,030 & 632,921 & 647,387\\ 
 \# Distinct Users & 143,481 & 202,907 & 30,939 & 136,847 & 386,757 & 303,643\\ 
 \bottomrule
\end{tabular} 
}
 \label{tab:ds-stats-extended}
\end{table*}
\section{Extended Analysis}
\label{sec:extend-obs}

\subsection{Degree-Related Patterns across News Topics}
Recall that we made two degree-related findings in Section \ref{sec:empirical-obs}, on how both \textit{degree} and \textit{co-engagement} of news articles closely relate to news veracity. To investigate if these observations are generalizable beyond political news (PolitiFact) and celebrity news (GossipCop), we extend our analysis to four additional news datasets covering three additional topics, namely three datasets from the MC-Fake benchmark \cite{min2022divide} on different topics (Syria War, Health and Covid-19) and the FANG dataset \cite{nguyen2020fang} (contains news articles about political events and influential rumor events). 

As our observations are based on social user engagement patterns, we focus on the news instances with social user engagements, and filter the instances without any user engagement. More specifically, we record social user engagements in terms of source tweets reposting news articles and their retweets, and collect the corresponding user IDs. The descriptive statistics of the datasets are summarized in Table \ref{tab:ds-stats-extended}.

\begin{figure*}[ht]
    \centering
    \includegraphics[width=0.9\textwidth]{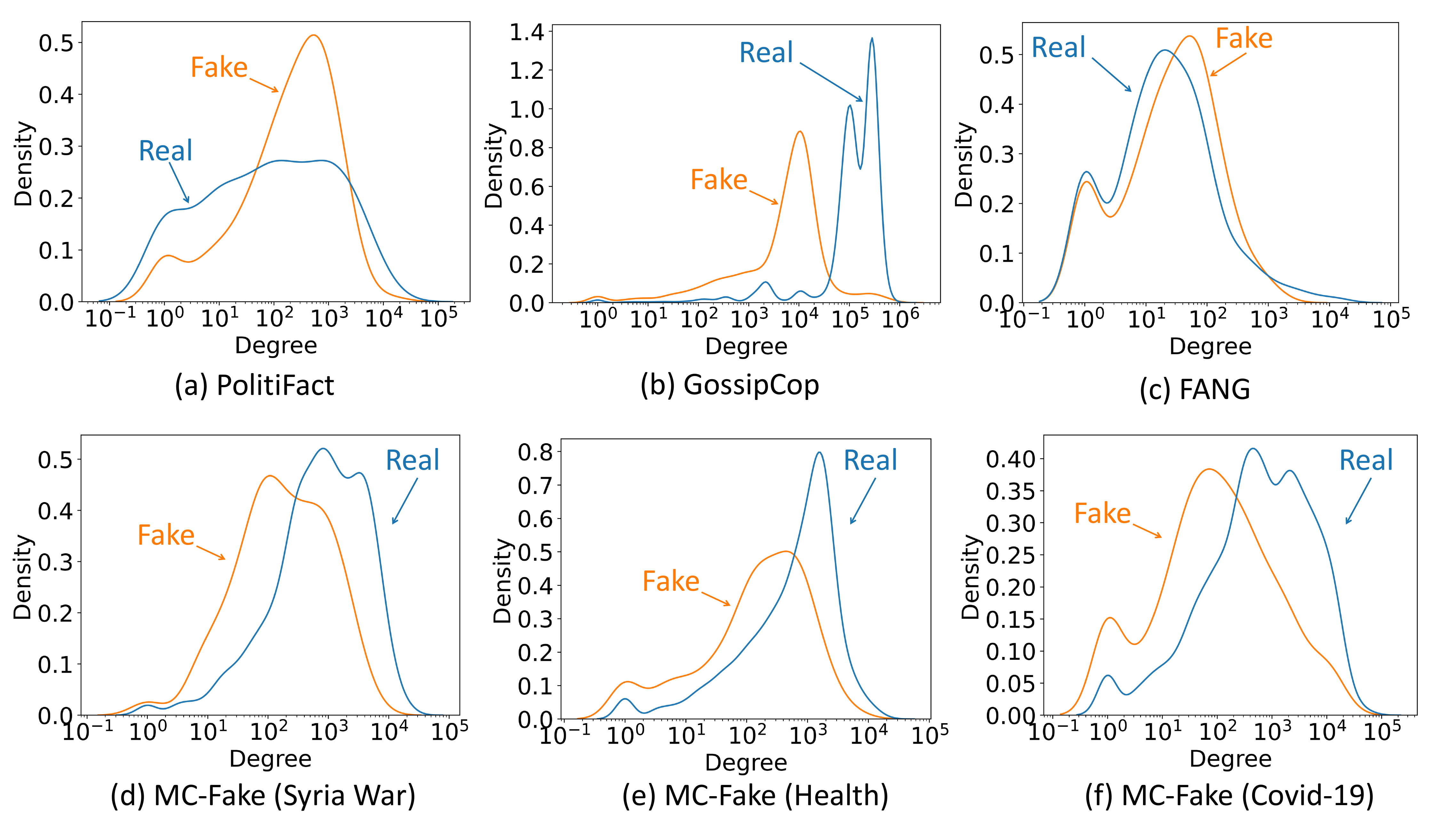}
    \caption{KDE plot of node degree distributions on the news engagement graph. }
    \label{fig:deg_kde_extend}
\end{figure*}

\begin{figure*}[ht]
    \centering
    \includegraphics[width=0.95\textwidth]{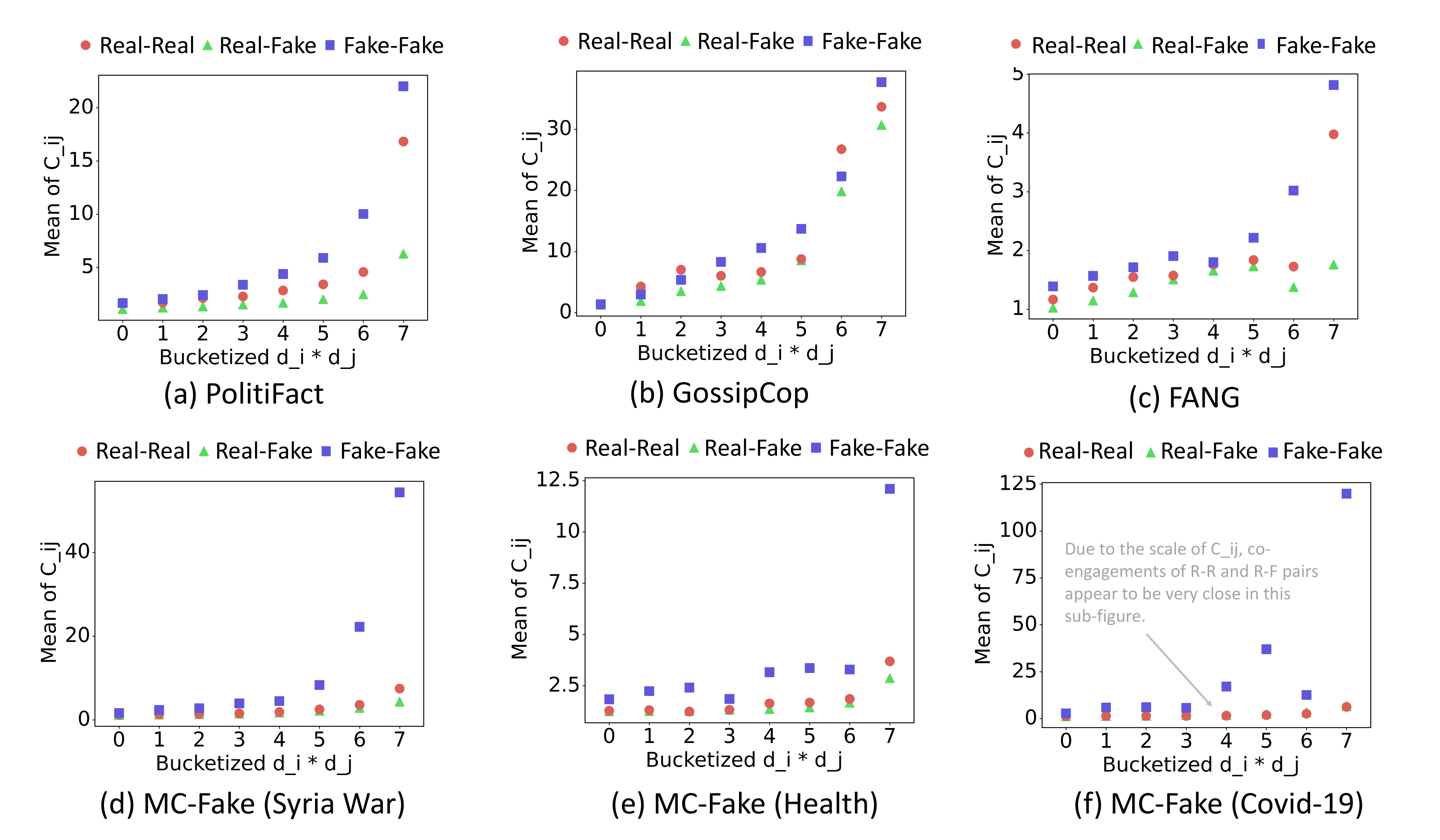}
    \caption{News co-engagement patterns of news article pairs. Edges are grouped based on the connected articles' veracity labels. }
    \label{fig:co_engage_extend}
\end{figure*}

Following the same procedure of plotting Figure \ref{fig:degree-cred} and Figure \ref{fig:co-engagement} in Section \ref{sec:empirical-obs}, we visualize the node degree distributions of real and fake news via KDE plots in Figure \ref{fig:deg_kde_extend}, and present the co-engagement patterns of news article pairs in Figure \ref{fig:co_engage_extend}. The plots are consistent with our two observations in that \textit{\textbf{(1)}} the degree distributions of nodes representing fake and real news articles exhibit a clear difference; and \textit{\textbf{(2)}} given the degrees, edges connecting fake and real news typically have the lowest co-engagement, whereas edges connecting fake news pairs typically have the highest co-engagement. This validates that our observed patterns are widely applicable to news of different topics, and demonstrates promising potential of applying veracity-related co-engagement and degree patterns to refine social graphs that involve news of varying topics.

\subsection{Discussion on Empirical Findings}

In this subsection, we discuss the probable reasons leading to our Observation \ref{obs_cij} (Section \ref{sec:empirical-obs}) on veracity-related patterns between co-engagement and degrees, which forms the key motivation of DECOR. Recall that Observation \ref{obs_cij} is two-fold: \textbf{(A)} Real-Fake news article pairs have the lowest co-engagement given the degrees; and \textbf{(B)} Fake-Fake pairs have higher co-engagement than real-real pairs given the degrees.

We find that both \textbf{(A)} and \textbf{(B)} closely relate with the confirmation bias theory \cite{nickerson1998confirm}, which states that users tend to seek and interpret evidence that upholds their existing beliefs, so as to gain confidence in their biased views. 

In terms of  \textbf{(A)}, as social media platforms foster echo chambers \cite{garimella18political} that insulate users from opposing viewpoints, social users tend to repeatedly engage in spreading news articles on certain topics with similar veracity. Hence, social users are less likely to share interest in two news articles of different veracity types (i.e., Real-Fake pairs), which accounts for lower co-engagement than Fake-Fake and Real-Real pairs with the same veracity type.

The underlying phenomena for \textbf{(B)} may vary, as the effects of confirmation bias can be manifested in different forms under different topics. Under topics such as politics \cite{vicario19polarization} or Covid-19 \cite{modgil21confirm}, social media platforms can induce opinion polarization, where the user's attention is highly segregated on a set of certain opinions. In terms of celebrity gossip, certain groups of social media users can engage in boundary coordination to gain control over the information \cite{mcnealy19tea}. These phenomena result in sharp community structures, which can be quantified via increased co-engagements.

In conclusion, our second empirical observation can be partially explained by multiple phenomena, and the underlying phenomena can differ across different topics. The benefits of our proposed DECOR framework (Section \ref{sec:decor}) is that through incorporating a learnable degree correction mechanism, the model is able to recognize the complex veracity-related degree patterns in a more flexible manner. Hence, DECOR facilitates effective detection of news articles without loss of computational efficiency.

\end{document}